# Forecasting the Remittances of the Overseas Filipino Workers in the Philippines

**Merry Christ E. Manayaga and Roel F. Ceballos**

Department of Mathematics and Statistics
College of Arts and Sciences
University of Southeastern Philippines
Bo. Obrero, Davao City, Philippines 8000
Email : manayagamerry@gmail.com
roel.ceballos@usep.edu.ph

**ABSTRACT**

*This study aims to find a Box-Jenkins time series model for the monthly OFW's remittance in the Philippines. Forecasts of OFW's remittance for the years 2018 and 2019 will be generated using the appropriate time series model. The data were retrieved from the official website of Bangko Sentral ng Pilipinas. There are 108 observations, 96 of which were used in model building and the remaining 12 observations were used in forecast evaluation. ACF and PACF were used to examine the stationarity of the series. Augmented Dickey Fuller test was used to confirm the stationarity of the series. The data was found to have a seasonal component, thus, seasonality has been considered in the final model which is $SARIMA\,(2,1,0) \times (0,0,2)_{12}$. There are no significant spikes in the ACF and PACF of residuals of the final model and the L-jung Box Q* test confirms further that the residuals of the model are uncorrelated. Also, based on the result of the Shapiro-Wilk test for the forecast errors, the forecast errors can be considered a Gaussian white noise. Considering the results of diagnostic checking and forecast evaluation, $SARIMA\,(2,1,0) \times (0,0,2)_{12}$ is an appropriate model for the series. All necessary computations were done using the R statistical software.*

**Journal of Economic Literature (JEL) Classification Number:** C53, F24

## 1. INTRODUCTION

Remittance is the money sent by migrant workers to their home countries. It is either sent through formal and informal channels. In some countries, remittances make up a decent portion of their GDP (Mohapatra and Ratha, 2010). Remittances play a huge role in the economy of developing countries like the Philippines for it can help alleviate poverty and create a stable cash flow and circulation in the economy.

Forecasting the OFW remittance is a necessary tool for analysing the impact of global financial crisis on the economy of developing countries. Furthermore, forecast values generated by a reliable model can be used to aid economic managers and policy makers to develop efficient plans and programs (Mohapatra and Ratha, 2010).

The aim of this research is to develop an appropriate model that can describe and forecast the OFW remittance using the Box-Jenkins Method.

## 2. METHODS

Montgomery, Jennings and Kulahci (2008), proposed the three iterative steps for Box-Jenkins forecasting method. It starts with Model Identification then followed by Model Estimation and Diagnostic Checking. An additional step called Model Evaluation is also suggested by several authors in order to assess the validity of the model. Thus, putting all these steps together, the resulting procedure is as follows:

- **Model Identification**

  The first thing to consider in forecasting is to determine if the series is stationary by checking if the mean and variance are stable. Two approaches can be used to check the stationarity of the series. These are by examining Autocorrelation Function (ACF) plots and by performing the Augmented Dickey Fuller (ADF) test. If the series is not stationary, transformation using one or the combination of differencing and Logarithmic Transformation can be done to achieve stationarity. Once the data is stationary, tentative or candidate models are identified using the ACF and Partial Autocorrelation Function (PACF) plots of the stationary series. Among the tentative models, the model with the least Akaike Information Criterion (AIC) value is chosen as the final model.

- **Model Estimation**

  The parameters of final model are then estimated using the combination of Conditional Sum of Squares and the method of Maximum Likelihood. The final model is considered good when the estimates of its parameters are all significant (p-value<0.05).

- **Diagnostic Checking**

  The diagnostic checking is done to assess if the model fits the data. The procedure for the diagnostic checking are as follows:
  
  a. Plots of the Residuals of the model were analysed to determine if the model fits the data. The model is considered to have a good fit if there are no visible patterns in the residual plots.
  
  b. The ACF and PACF plots of the residuals must be a white noise, that is, there should be no significant spikes in both the ACF and PACF plots.
  
  c. The Ljung-Box test was used as a formal test to determine if the model is a good fit to the data (Ljung, 1978)

- **Model Validation/Forecast Evaluation**

  Model evaluation is used to evaluate forecast accuracy of the model. The one-step ahead forecast was generated using the final model. The forecast errors are obtained by taking the difference between the 10 data points that were not part of the model building (actual values) and the one-step ahead forecasts. If there are no significant spikes in the ACF and PACF plots of the forecast errors, then it is considered a white noise. The Shapiro-Wilk test (Shapiro, 1965) was used to test for the normality of the forecast errors. The Mean Absolute Percentage Error (MAPE) was used to determine the accuracy of the forecast of the model (Hyndman, 2006). The MAPE should be less than 5% so that the model is considered accurate.

## 3. RESULTS

Figure 1 shows the time series plot of monthly remittance from sea-based and land-based OFW. There were 96 data points from the year 2009 to 2016. The highest point recorded was the latest remittance which is in December 2016. The remittance in January 2009 was the lowest in the given data. An upward trend is obviously present when observing the time series plot. Also, seasonality is observed specifically from 2009 to 2014. There is also a slight problem with the variance of the series therefore, Box-Cox transformation should be done on the data.

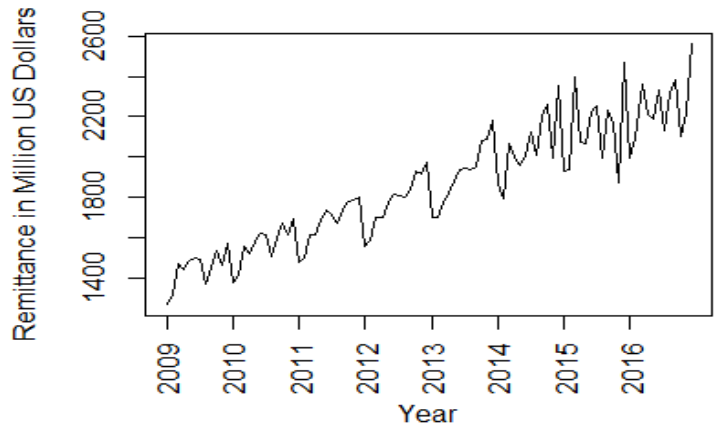

Figure 1. Monthly OFW Remittance in the Philippines

.
The Augmented Dickey-Fuller test was used to determine if the transformed monthly OFW Remittance is stationary. Table 1 shows the result of the ADF test. The result shows that the p-value is large which means that the null hypothesis cannot be rejected. It is concluded that the transformed series is still nonstationary.

Table 1. Test for Stationary of the Transformed Series

| Dickey-Fuller Value | p-value |
| --- | --- |
| 1.2103 | 0.9395 |

The ACF plot of the transformed series also shows that the series is nonstationary. The ACF in Figure 3 is slowly decaying which is the characteristic of a nonstationary series. To achieve stationarity, first differencing is applied to the transformed series.

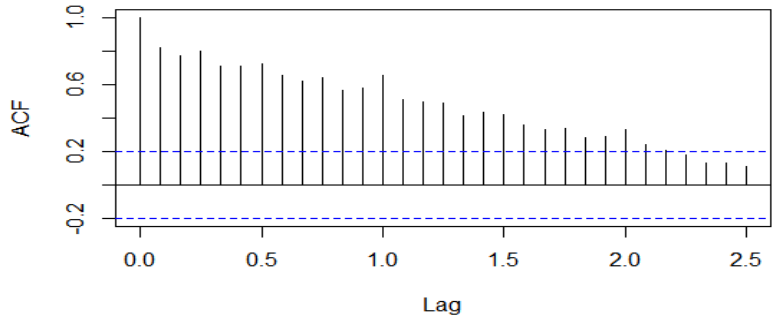

Figure 3. Plot of the Transformed series

Figure 4 shows the plot of the differenced data. The differenced data seems to be stationary since no trend is visible. ADF test was done to test if stationarity is now achieved since the series has already been differenced.

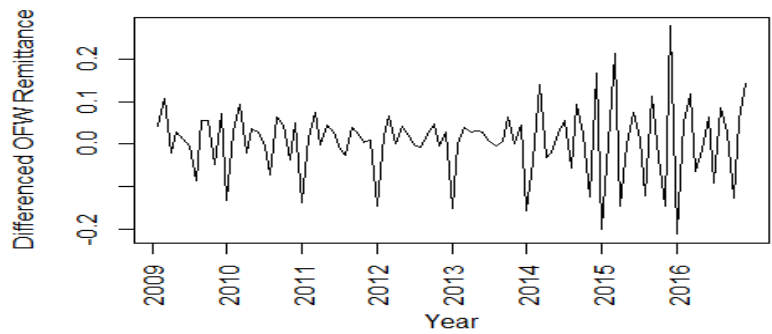

Figure 4. Plot of the Differenced and Transformed Series

Table 2 shows that the p-value of the ADF test for the differenced data is less than 0.05. Therefore, the null hypothesis is rejected and it is concluded that the differenced data is stationary.

Table 2. Test for stationary of the differenced series

| Dickey-Fuller Value | p-value |
| --- | --- |
| -13.79 | <0.01 |

Figure 5 shows that the ACF of the differenced series have improved since it shows a damped sinusoidal pattern. The slowly decaying pattern is no longer present**.**

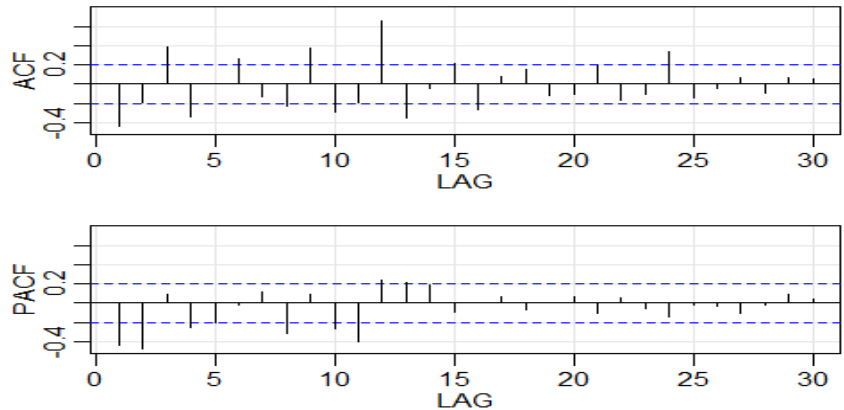

Figure 5. ACF and PACF of differenced series

There are several interpretations that can be drawn from the ACF and PACF of the differenced data (see Figure 5). For the non-seasonal aspect of the model, the spikes at lag 1 of the ACF and the PACF is considered. The spike at lag 2 of the PACF is also considered. For the seasonal part, the spikes at lag 24 of the ACF will be considered cut off while the spike at lag 12 of the PACF is also considered cut off which suggests two components of seasonal moving average or $SMA(2)$ and one seasonal autoregressive or $SAR(1)$. Table 3 shows the tentative models that are considered and their respective Akaike's Information Criteria values. It shows that the model $SARIMA\,(2,1,0) \times (1,0,0)_{12}$ has the lowest Akaike Information Criterion of -327.35. However, the residual plots of the said model show a funnelling behaviour and there are few spikes in the ACF and PACF of the residual. The next model with the smallest AIC, $SARIMA\,(1,1,1) \times (1,0,0)_{12}$ has also a nonsignificant coefficient. Thus, the chosen model for this series is $SARIMA\,(2,1,0) \times (0,0,2)_{12}$.

Table 3. The Tentative ARIMA models for OFW Remittance

| Tentative Models | AIC |
|---|---|
| $SARIMA\,(1,1,1) \times (1,0,0)_{12}$ | -323.07 |
| $SARIMA\,(2,1,0) \times (1,0,0)_{12}$ | -327.35 |
| $SARIMA\,(1,1,1) \times (0,0,2)_{12}$ | -314.86 |
| $SARIMA\,(2,1,0) \times (0,0,2)_{12}$ | -315.01 |

Table 4 shows the estimated coefficient, standard error, the z-value as well as the p-value of the coefficient of the model. The p-value of the MA parameter and the AR parameter for both non-seasonal and seasonal components are significant at α=0.01. It is concluded that the parameters are significantly different from zero.

Table 4. Model Estimates

| Tentative Models | AR(1) | AR(2) | SMA(1) | SMA(2) |
|---|---|---|---|---|
| Estimate | 0.6877 | -0.4831 | 0.9972 | 0.4131 |
| Standard error | 0.0980 | 0.0937 | 0.1207 | 0.1147 |
| z-value | -7.0185 | -5.1548 | 8.2612 | 3.6029 |
| p-value | <0.01 | <0.01 | <0.01 | <0.01 |

To determine if the chosen model is adequate, diagnostic check was done. The residual analysis is presented in Figure 6. From the first plot (Residual vs. Time), it is evident that there is no correlation among the residuals since the residuals are randomly positioned about the area of the plot. Also, there are no obvious patterns or systematic trends in the second plot (Residual vs. Fitted). The residual does not drift far away from the theoretical line as shown in the third plot (qq plot), which implies that the normality assumption has been met.

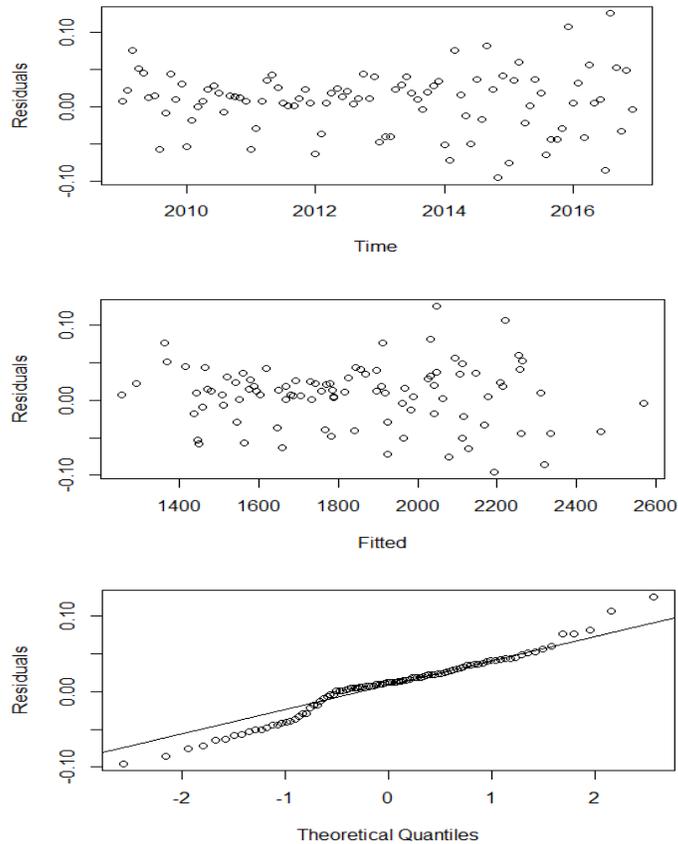

Figure 6. Residual Plots for Final Model

Figure 7 shows the ACF and PACF plot of residuals. Although there are few lags that touch the limit, all of the lags are within the acceptable limit. This means that the residuals are uncorrelated.

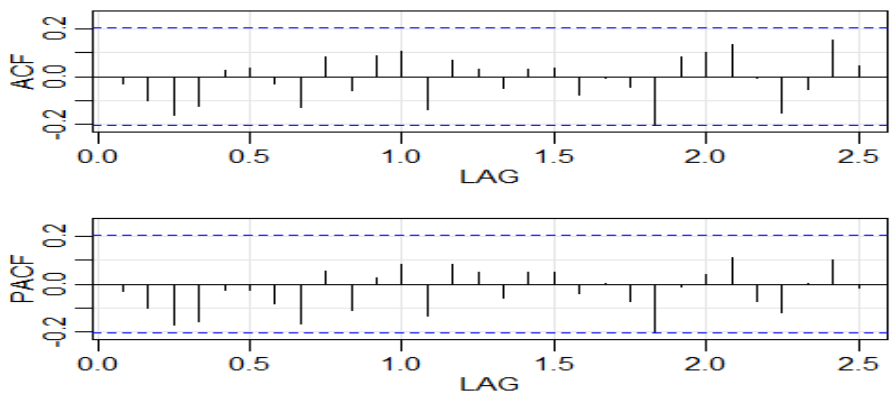

Figure 7. ACF and PACF Plots of Residual

Ljung-Box Q* Test is conducted to formally test if the residuals are uncorrelated and the model does not exhibit lack of fit. Table 5 shows the result of the test applied to the model's residual. Since the p-value is large, the null hypothesis cannot be rejected. It is concluded that the model does not exhibit lack of fit.

Table 5. Ljung-Box Q* Test

| Ljung-Box Q* Test Statistic | p-value |
|---|---|
| 22.443 | 0.317 |

Model validation is done to make sure that the model performs well in forecasting. Table 6 shows the actual value, one-step ahead forecast and the forecast error for the month of January 2017 up to December 2017. Forecast error is acquired by simply subtracting the one step ahead forecast from the actual value.

Table 6. One Step Ahead Forecasts and Forecast Errors

| Month | Year | Actual Value | One Step Ahead Forecast | Forecast Error |
|---|---|---|---|---|
| Jan | 2017 | 2168.700 | 2145.530 | 23.170 |
| Feb | 2017 | 2169.241 | 2236.218 | -66.977 |
| Mar | 2017 | 2615.216 | 2457.323 | 157.893 |
| Apr | 2017 | 2082.618 | 2337.417 | -254.799 |
| May | 2017 | 2309.758 | 2306.372 | 3.386 |
| Jun | 2017 | 2467.073 | 2398.420 | 68.653 |
| Jul | 2017 | 2282.731 | 2237.650 | 45.081 |
| Aug | 2017 | 2499.483 | 2408.218 | 91.265 |
| Sep | 2017 | 2186.091 | 2475.588 | -289.497 |
| Oct | 2017 | 2275.151 | 2214.368 | 60.783 |
| Nov | 2017 | 2262.313 | 2289.175 | -26.862 |
| Dec | 2017 | 2741.425 | 2580.624 | 160.801 |

*Note: MAPE = 4.1%*

It is ideal that the forecast errors behave like a Gaussian white noise. The forecast errors are considered white noise if the ACF and PACF are within the upper and lower bound limits. Figure 10 shows that the ACF and PACF are both within the limits. There are no significant spikes in the both the ACF and PACF plots. Therefore, the forecast errors behave like a white noise process. Furthermore, the mean absolute percentage error was found to be 4.1% which implies that the one-step ahead forecast is accurate.

Figure 8. ACF and PACF Plots of Forecast errors

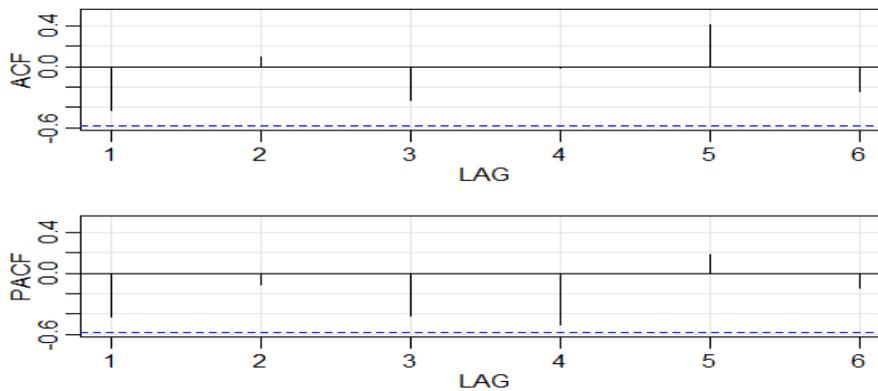

Furthermore, the forecast errors should be normally distributed so that it can be considered Gaussian white noise. Table 7 shows the result of the Shapiro-Wilk test. Since the p-value is not significant at α=0.01, the null hypothesis cannot be rejected. It is concluded that the forecast error is normally distributed. The forecast errors behave like a Gaussian white noise in this case.

Table 7. Test for Normality of Forecast Errors

| Shapiro-Wilk Test Statistics | p-value |
|---|---|
| 0.8723 | 0.06991 |

Table 8 shows the forecasted values of OFW remittance for the year 2018 to 2019 and their respective intervals at 95% confidence coefficient. In forecasting the OFW remittance using the model ARIMA (2, 1, 0) × (0, 0, 2)12, it is predicted that the highest OFW remittance would be on March of 2018 with a total of 2760.508 million US Dollars. All forecast values are within the lower and upper limits.

Table 8. Forecasted OFW Remittance

| Month | Year | Forecast Values | Lower 95% Confidence Interval | Upper 95% Confidence Interval |
|---|---|---|---|---|
| Jan | 2018 | 2380.304 | 2193.853 | 2582.602 |
| Feb | 2018 | 2213.33 | 2032.048 | 2410.786 |
| Mar | 2018 | 2770.618 | 2534.845 | 3028.32 |
| Apr | 2018 | 2176.305 | 1962.945 | 2412.855 |
| May | 2018 | 2384.249 | 2139.008 | 2657.608 |
| Jun | 2018 | 2532.455 | 2260.764 | 2836.796 |
| Jul | 2018 | 2467.976 | 2186.301 | 2785.94 |
| Aug | 2018 | 2496.712 | 2200.005 | 2833.436 |
| Sep | 2018 | 2222.445 | 1948.494 | 2534.912 |
| Oct | 2018 | 2464.478 | 2148.359 | 2827.11 |
| Nov | 2018 | 2288.519 | 1985.13 | 2638.275 |
| Dec | 2018 | 2702.518 | 2333.143 | 3130.372 |
| Jan | 2019 | 2484.152 | 2055.532 | 3002.147 |
| Feb | 2019 | 2339.944 | 1916.642 | 2856.735 |
| Mar | 2019 | 2672.583 | 2168.622 | 3293.659 |
| Apr | 2019 | 2364.126 | 1883.113 | 2968.006 |
| May | 2019 | 2444.147 | 1925.825 | 3101.973 |
| Jun | 2019 | 2537.122 | 1978.751 | 3253.056 |
| Jul | 2019 | 2530.946 | 1949.287 | 3286.169 |
| Aug | 2019 | 2460.591 | 1875.822 | 3227.656 |
| Sep | 2019 | 2367.957 | 1787.637 | 3136.666 |
| Oct | 2019 | 2508.453 | 1874.337 | 3357.1 |
| Nov | 2019 | 2376.63 | 1759.21 | 3210.743 |
| Dec | 2019 | 2570.511 | 1885.503 | 3504.385 |

## 4. CONCLUSION AND RECOMMENDATION

The following conclusions and recommendations are made as a result of the study.

- Remittances of OFW in the Philippines exhibits trend and seasonal components. The peak occurs on the month of December of every year.
- Furthermore, the forecast errors should be normally distributed so that it can be considered Gaussian white noise. Table 7 shows the result of the Shapiro-Wilk test.
- The appropriate model for the Remittances of OFW in the Philippines is

$$z_t = 0.6877z_{t-1} + 0.4831z_{t-2} + 0.9972\varepsilon_{t-12} + 0.4131\varepsilon_{t-24} + \varepsilon_t$$

- Based on the diagnostic checks and forecast evaluations the model fits the data and it can be used to forecast the monthly Remittances of the OFW in the Philippines.